\documentclass[12pt]{article}

\usepackage{cite}

\usepackage{graphicx}
\usepackage{amsmath}
\usepackage{amssymb}
\allowdisplaybreaks

\begin{document}

\baselineskip=17.5pt plus 0.2pt minus 0.1pt

\renewcommand{\theequation}{\arabic{equation}}
\renewcommand{\thefootnote}{\fnsymbol{footnote}}
\makeatletter
\def\CR{\nonumber \\}
\def\be{\begin{equation}}
\def\ee{\end{equation}}
\def\bea{\begin{eqnarray}}
\def\eea{\end{eqnarray}}
\def\bead{\be\begin{aligned}}
\def\eead{\end{aligned}\ee}
\def\eq#1{(\ref{#1})}
\def\la{\langle}
\def\ra{\rangle}
\def\hyp{\hbox{-}}
\def\ul{\underline}

\begin{titlepage}
\title{\hfill\parbox{4cm}{ \normalsize YITP-11-93}\\
\vspace{1cm} Canonical tensor models with local time}
\author{Naoki {\sc Sasakura}
\thanks{\tt sasakura@yukawa.kyoto-u.ac.jp}
\\[15pt]
{\it Yukawa Institute for Theoretical Physics, Kyoto University,}\\
{\it Kyoto 606-8502, Japan}}
\date{}
\maketitle
\thispagestyle{empty}
\begin{abstract}
\normalsize
It is an intriguing question how {\it local} time can be introduced in the emergent picture of spacetime.
In this paper, this problem is discussed in the context of tensor models.
To consistently incorporate {\it local} time into tensor models, 
a rank-three tensor model with first class constraints in Hamilton formalism
is presented.
In the limit of usual continuous spaces, the algebra of constraints reproduces that of
general relativity in Hamilton formalism. 
While the momentum constraints can be realized rather easily by the symmetry of 
the tensor models, the form of the Hamiltonian constraints  
is strongly limited by the condition of the closure of the whole constraint algebra.
Thus the Hamiltonian constraints have been determined on the assumption that
they are local and at most cubic in canonical variables.   
The form of the Hamiltonian constraints has similarity with the Hamiltonian in the $c<1$ string
field theory, 
but it seems impossible to realize such a constraint algebras 
in the framework of vector or matrix models.
Instead these models are rather useful as matter theories coupled with the tensor model. 
In this sense, a three-index tensor is the minimum-rank dynamical 
variable necessary to describe gravity in terms of tensor models. 
\end{abstract}
\end{titlepage}

\section{Introduction}
\label{sec:intro}

Tensor models have originally been introduced \cite{Ambjorn:1990ge,Sasakura:1990fs,Godfrey:1990dt}
to describe the simplicial quantum gravity in general dimensions higher than two,
with the hope to extend the success of the matrix models in the study of the two-dimensional
simplicial quantum gravity.
Tensor models have later been extended to describe the spin foam and loop quantum gravities by
considering Lie-group valued indices \cite{Boulatov:1992vp,Ooguri:1992eb,DePietri:1999bx}.
These models with group indices, called group field theory \cite{Oriti:2011jm}, 
are actively studied with various interesting recent progress \cite{Gurau:2011sk,
Gurau:2011xp,Bonzom:2011ev,Benedetti:2011nn,
Baratin:2011tx,Gurau:2011tj,Bonzom:2011zz,Livine:2011yb,Carrozza:2011jn,Gurau:2011xq,
Baratin:2011tg,Gurau:2010ba,Geloun:2010vj,Gurau:2010nd,Gurau:2009tz,Gurau:2009tw}.
A key issue is the emergence of a new kind of tensor models,
called colored tensor models \cite{Gurau:2011xp}, which have 
more intimate topological correspondence to simplicial manifolds than the original versions.
There has also been a systematic study of tensor models in semi-classical 
approximations \cite{Sasakura:2010rb,Sasakura:2009dk,Sasakura:2009hs,Sasakura:2008pe,
Sasakura:2007ud,Sasakura:2007sv,Sasakura:2006pq,Sasakura:2005gv}
under the interpretation of the rank-three tensor models as theory of dynamical fuzzy spaces
\cite{Sasakura:2011ma,Sasakura:2005js}, 
and the emergence of Euclidean general relativity on emergent spaces has been observed 
\cite{Sasakura:2008pe,Sasakura:2009hs}.

So far the study of tensor models has basically been limited to the cases of Euclidean signature.
In the standard field theories on flat spaces, after setting up appropriate causal structures,
the computations in Minkowski signature can
be obtained from those in Euclidean signature by means of analytic continuation.
However, it is generally not clear how one may extend the standard procedures such as
Wick rotation to the quantum gravitational situation with fluctuating geometries. 
Moreover, the study of the causal dynamical triangulation \cite{Ambjorn:2010rx}
indicates that the dynamics of quantum gravity in Minkowski signature may be
substantially different from that in Euclidean signature.
The main purpose of this paper is to discuss how to incorporate time in the framework 
of tensor models.

The advent and correctness of the theory of relativity have established that
time is not an absolute entity but is rather a relative quantity measured by 
physical phenomena dubbed as a ``clock".
Since a ``clock" is a local object due to the speed limit of light, 
the definition of time is necessarily local
and is generally dependent on how the system of ``clocks" is organized.
The principle that physical phenomena themselves should not depend on 
this kind of ambiguity of defining local time provides strong constraints on 
possible forms of consistent theories of nature.

In the Lagrangian formalism, this ambiguity can well be incorporated
by imposing the invariance of theories under the general coordinate transformations of 
spacetime coordinates. In the rank-three tensor models for instance, however, the only 
dynamical variable is a three-index tensor, $M_{abc}\ (a,b,c=1,2,\ldots,N)$, 
and there are neither space nor locality built in the framework:
a space and its locality are emergent phenomena \cite{Sindoni:2011ej}.
Therefore one would have difficulties in introducing local time into the tensor models and 
imposing the constraints coming from the principle mentioned above.
One would try to introduce time $t$ simply as an argument of the
tensor as $M_{abc}(t)$, and impose the invariance of the models
under reparametrization $t'(t)$. 
However, this way of introducing time will necessarily allow 
an entity of a non-local global time to exist when a space is emergent. 
Then emergent field theories on the emergent space
will not generally satisfy the above principle; emergent field theories on an emergent
flat space will seriously violate Lorentz symmetry, that is rather strongly constrained 
experimentally and theoretically  
\cite{Polchinski:2011za,Gambini:2011nx,Collins:2004bp}.
Another option to try would be to introduce time for each index of the tensor as 
$M_{a\, t_a, b \,t_b, c \, t_c}$. This option seems to have an appealing feature
concerning locality of time. However, the contraction of indices of the tensor such as 
$\sum_{abc} \int dt_a dt_b dt_c\, M_{a\, t_a, b \,t_b, c \, t_c}M_{a\, t_a, b \,t_b, c \, t_c}$
will introduce multiple integrals over times into the Lagrangian formalism.
Therefore this option seems to require extension of the Lagrangian formalism 
for multiple time parameters in advance, before applying it to the tensor models. 

The above lack of good guiding principles for introducing time to the tensor variable 
would require us to abandon starting with Lagrangian formalism for the purpose.
There exists another formalism of mechanics, Hamilton formalism, 
in which time is rather intimately related to dynamical evolution of a system. 
Of course, the two formulations of 
mechanics are equivalent (at least classically), but in the present confusing situation 
on local time, the latter formalism is superior to the former one, because 
one does not have to know in advance how time is represented in the dynamical variable.
If necessary, once the Hamilton formalism of the tensor models is consistently obtained, 
one would also be able to obtain the corresponding Lagrangian formalism.
    
In Hamilton formalism of general relativity, the consistency of dynamics under ambiguous 
choices of local time is guaranteed by a set of first class constraints, 
which are the generators of the local coordinate transformations containing the time direction.
In this paper, the set of constraints of general relativity will be rewritten
in terms of the dynamical variables of tensor models, 
and a rank-three tensor model with first class constraints in Hamilton formalism will be
presented. This should be equivalent to introducing local time in tensor models.

The canonical formulations of discrete models of gravity have been discussed in 
previous literatures 
\cite{Piran:1985ke,Friedman:1986uh,Gambini:2005vn,Dittrich:2011ke,Bahr:2011xs}. 
An important difference of the present work from the previous approaches
is that the spatial diffeomorphism is exactly represented by the symmetry of the tensor model.
Therefore the gauge freedom of spatial diffeomorphism is exactly incorporated by first class
momentum constraints of the tensor model. 
Then Hamiltonian constraints will be determined by the condition of the closure
of the whole set of first class constraints. 
It is also peculiar that time is just a continuous variable unlike some previous approaches
\cite{Gambini:2005vn,Dittrich:2011ke}. 
Thus the standard Dirac procedure is applicable, and 
a rank-three tensor model with local time will be formulated in terms of
the standard Hamilton formalism with a set of first class constraints.

This paper is organized as follows.
In Section \ref{sec:gr}, 
the general relativity in Hamilton formalism 
is summarized for the discussions in this paper.
In Section \ref{sec:reviewtensor}, the rank-three tensor models are briefly overviewed.
The limit of usual continous spaces in the tensor models is explained. 
In Section \ref{sec:vector}, an algebra expressed in terms of the canonical variables 
of a vector model is considered, and it is shown that  
the algebra of the first class constraints of general relativity
can be reproduced in the limit of usual continuous spaces.
In Section \ref{sec:difficulty}, however, some difficulties in the realization in Section \ref{sec:vector}
are pointed out, and the necessity of a three-index tensor is argued.
In Section \ref{sec:tensor}, a rank-three tensor model with first class constraints in Hamilton formalism 
is presented. The momentum constraints represent the kinematical
symmetry of the rank-three tensor model, and
contain the spatial diffeomorphism symmetry in the limit of usual continuous spaces. 
The Hamiltonian constraints are determined by the closure of the algebra of the whole first class constraints
under the assumption that they are at most cubic and respect locality.   
It turns out to be necessary to break the time-reversal symmetry.
In Section \ref{sec:matter}, it is shown that matter degrees of freedom coupled to the rank-three tensor
model can be added to the system without destroying the algebraic structure of the first class constraints.
The matter degrees of freedom can be given by any rank tensors. 
Section \ref{sec:sum} is devoted to summary, discussions and future prospects.

\section{The first class constraints from general relativity}
\label{sec:gr}
The ADM formulation \cite{Arnowitt:1962hi} of general relativity leads to the following algebra of first class 
constraints\footnote{The trivial ones with the conjugate momentums  $\pi_N,\pi_{N_i}$ of the lapse and shift variables are 
omitted.} \cite{DeWitt:1967yk,1994LNP...434.....E},
\begin{align}
\label{eq:const1}
\{ {\cal H}(x), {\cal H}(x')\}&=\epsilon\left( {\cal H}^i(x) \delta_i(x,x')-{\cal H}^i(x')\delta_i(x',x)
\right),\\
\label{eq:const2}
\{ {\cal H}_i(x), {\cal H}(x')\}&={\cal H}(x) \delta_i(x,x'),\\
\label{eq:const3}
\{ {\cal H}_i(x), {\cal H}_j(x')\}&={\cal H}_i(x') \delta_j(x,x')+{\cal H}_j(x)\delta_i(x,x'),
\end{align}
where $\{\ ,\ \}$ denotes Poisson bracket, and $\delta_i(x,x')$ denotes the derivative of 
the delta function with respect to $x^i$. 
The signature $\epsilon$ takes $\epsilon=1$ and $\epsilon=-1$ 
for Minkowski and Euclidean signatures, respectively.
${\cal H}(x)$ and ${\cal H}_i(x)$ are the super-Hamiltonian and the super-momentums, respectively,
and they are explicitly given by
\begin{align}
\label{eq:hamiltonconst}
{\cal H}(x)&=\frac{16 \pi G}{\sqrt{g}}\left(\pi_{ij}\pi^{ij}-\frac{1}{2}\left(\pi_i{}^i\right)^2 \right)-
\frac{\sqrt{g}}{16\pi G} \left(R^{(3)}-2 \Lambda\right),\\
{\cal H}_i(x)&=-2 D_j\pi_i{}^j,
\end{align}
where  $\pi_{ij}(x)$ is the conjugate momentum to the spatial geometry $g_{ij}(x)$, and $G$ and $\Lambda$
are the gravitational and cosmological constants, respectively.
The algebra \eq{eq:const1}, \eq{eq:const2} and \eq{eq:const3} can be equivalently 
expressed as
\begin{align}
\label{eq:grahh}
\{H(v),H(w)\}&=\epsilon D(v\partial^iw-w\partial^iv),\\
\label{eq:gradh}
\{D(v^i),H(w)\}&=H(v^i\partial_iw),\\
\label{eq:gradd}
\{D(v^i),D(w^i)\}&=D(v^j\partial_jw^i-w^j\partial_jv^i),
\end{align}
where $H(v)$ and $D(w^i)$ are defined by
\begin{align}
H(v)&\equiv \int dx\, v(x) {\cal H}(x),\\
D(v^i)&\equiv \int dx\, v^i(x) {\cal H}_i(x),
\end{align}
with $v(x)$ and $v^i(x)$ independent of the canonical variables $g_{ij}(x)$ and $\pi_{ij}(x)$.

The fundamental roles of the first class constraint algebra \eq{eq:const1}, \eq{eq:const2} and \eq{eq:const3}
in geometrodynamics have been discussed in \cite{Hojman:1976vp}.
An important feature is that the right-hand side of \eq{eq:const1} contains the inverse spatial metric $g^{ij}(x)$
to raise the index of ${\cal H}_i(x)$ to ${\cal H}^i(x)$. 
Therefore the constraint algebra has structure functions depending on the canonical variable,
and is not a Lie algebra with constant structure constants. 
To reconcile the constraint algebra with the spacetime diffeomorphism, 
the on-shell conditions ${\cal H}={\cal H}_i=0$ must be imposed.  
In addition,
under some (reasonable) assumptions, the form of the super-Hamiltonian has 
been shown to be given uniquely by the form \eq{eq:hamiltonconst}.
It is also argued that the constraint algebra does not change by adding 
matters with non-derivative couplings with gravity. 

\section{A brief overview of tensor models}
\label{sec:reviewtensor}
This section will give a brief overview of the rank-three tensor models (with no time) 
to prepare for the discussions in the following sections.
As explained in Section \ref{sec:intro}, tensor models have
some variations with distinct interpretations.
This paper deals with the rank-three tensor models, which have 
a three-index tensor as their only dynamical variable.
The rank-three tensor models can be regarded as theory of dynamical fuzzy spaces 
\cite{Sasakura:2011ma,Sasakura:2005js}. 
This interpretation of the tensor models is especially convenient to understand 
the semi-classical behavior of the tensor models: classical solutions are 
regarded as background fuzzy spaces which approximate continuum spaces, and 
the perturbations around the classical solutions are interpreted by effective field theories
on the spaces. In fact, various classical solutions and the perturbations around them are 
studied to show the phenomena of emergent Euclidean general relativity on emergent spaces 
\cite{Sasakura:2010rb,Sasakura:2009dk,Sasakura:2009hs,Sasakura:2008pe,
Sasakura:2007ud,Sasakura:2007sv,Sasakura:2006pq,Sasakura:2005gv}.
It is especially noteworthy that, as theory of dynamical fuzzy spaces, the rank-three tensor models
can deal with any dimensional spaces, unlike the original proposals 
\cite{Ambjorn:1990ge,Sasakura:1990fs,Godfrey:1990dt},
in which the ranks of tensors are related with dimensions.

Let me denote the three-index dynamical tensor by $M_{abc}$. 
The tensor is assumed to satisfy the generalized Hermiticity condition 
\cite{Ambjorn:1990ge,Sasakura:1990fs,Godfrey:1990dt},
\be
M_{abc}=M_{bca}=M_{cab}=M^*_{bac}=M^*_{acb}=M^*_{cba},
\label{eq:tengenher1}
\ee
where $^*$ denotes complex conjugation, and the indices run as $a,b,c=1,2,\ldots,N$.
Because of the generalized Hermiticity condition \eq{eq:tengenher1}, 
the symmetry which can be associated
to the rank-three tensor models is the orthogonal group $O(N)$, 
\be
M'_{abc}=O_a{}^{a'}O_b{}^{b'}O_c{}^{c'}M_{a'b'c'},\ \ \ O_a{}^b\in O(N),
\label{eq:tensorsym}
\ee
instead of a unitary group of an hermitian matrix model.

While a continuous manifold can be described by a coordinate system, a fuzzy space
is defined by an algebra of functions on it. 
The algebra of functions $\phi_a\ (a=1,2,\ldots,N)$ can be characterized by its structure constants
$C_{abc}$ as\footnote{Repeated indices are assumed to be summed over throughout this paper, 
unless otherwise stated.}
\be
\phi_a \phi_b= C_{abc} \phi_c.
\label{eq:fuzzyalg}
\ee
While a usual continuous space can be characterized
by a commutative and associative algebra of functions, a noncommutative 
space for instance can be characterized by a noncommutative associative algebra.
One may even consider a nonassociative algebra to define a nonassociative space
\cite{Ramgoolam:2001zx,Ramgoolam:2003cs,deMedeiros:2004wb,Sasai:2006ua}.

To relate fuzzy spaces to the configurations of the rank-three tensor models, 
let me introduce an inner product \cite{Sasakura:2011ma},
\be
\langle \phi_a | \phi_b \rangle=\delta_{ab},
\ee 
which is assumed to be bi-linear.
Now let me identify the structure constants with the dynamical variable of the tensor model,
\be
C_{abc}=M_{abc}.
\label{eq:CeqM}
\ee 
This physically means that 
the rank-three tensor models are interpreted as theory describing the dynamics of fuzzy spaces. 
The identification \eq{eq:CeqM} and the generalized Hermiticity condition \eq{eq:tengenher1} imply
the following cyclicity property on the algebraic structure,
\be
\langle \phi_a \phi_b | \phi_c \rangle=\langle \phi_a |\phi_b \phi_c \rangle =\langle \phi_b |\phi_c \phi_a \rangle,
\label{eq:algcon1}
\ee
and also the properties on complex conjugation, 
\begin{align}
\phi_a&=\phi_a^*,
\label{eq:algcon2}\\
(\phi_a \phi_b)^*&=\phi_b \phi_a.
\label{eq:algcon3}
\end{align}
These properties \eq{eq:algcon1}, \eq{eq:algcon2} and \eq{eq:algcon3} 
characterize the fuzzy spaces which can be associated with the configurations of the rank-three
tensor models. The fuzzy spaces with these properties have various interesting properties
concerning symmetries, uncertainties, and reduction procedures 
\cite{Sasakura:2011ma,Sasakura:2011nj,Sasakura:2011qg,Sasakura:2011talk}.

In the analysis of emergent Euclidean general relativity from the rank-three tensor models 
\cite{Sasakura:2010rb,Sasakura:2009dk,Sasakura:2009hs,Sasakura:2008pe,
Sasakura:2007ud,Sasakura:2007sv}, 
the following particular form of $M_{abc}$ with Gaussian functions has extremely been useful. 
The indices of functions $\phi_a$ are assumed to be 
given by the coordinates of a usual continuous $D$-dimensional space,
\be
a=x=(x^1,x^2,\ldots,x^D),\ \ \ x^i \in {\bf R},
\label{eq:contx}
\ee
and $M_{x_1x_2x_3}$ is assumed to be given by the following Gaussian form,  
\be
M^G_{x_1x_2x_3}=B\, g(x_1)^{1/4}g(x_2)^{1/4}g(x_3)^{1/4}
\exp \left[-\beta \left( d(x_1,x_2)^2+d(x_2,x_3)^2+d(x_3,x_1)^2\right)\right],
\label{eq:GaussM}
\ee
where $B$ and $\beta$ are positive constants. The fuzziness of the spaces is characterized 
by the parameter $\beta$. Here a metric tensor field $g_{ij}(x)$ is 
assumed to exist on the $D$-dimensional space, $g(x)=\hbox{Det}[ g_{ij}(x)]$, and 
$d(x_1,x_2)$ denotes the geodesic distance between two points $x_1$ and $x_2$. 
The form of \eq{eq:GaussM} respects the diffeomorphism symmetry, as $g(x)^{1/4}$
guarantees the diffeomorphism invariance of an index contraction:
$M^G_{xx_1x_2}M^G_{xx_3x_4}=\int dx \sqrt{g(x)}\cdots$.

The Gaussian configuration \eq{eq:GaussM} is merely an idealized working hypothesis
which singles out the modes corresponding to those of general relativity.
This hypothesis has very well explained the qualitative features of the results of the numerical 
analysis \cite{Sasakura:2009dk,Sasakura:2009hs,Sasakura:2008pe,
Sasakura:2007ud,Sasakura:2007sv}, which have shown the emergence of Euclidean general relativity
on emergent spaces.
The detailed values of $M_{abc}$ may be different from that, because the actual 
degrees of freedom of the rank-three tensor models are discrete and finite, 
while they are continuous and infinite for the continuum index \eq{eq:contx}. 
Also the simple Gaussian damping form is idealizing $M_{x_1x_2x_3}$ which is 
locally distributed
with respect to the relative locations of $x_i$. 
Thus the Gaussian configuration \eq{eq:GaussM} should be regarded as an infrared effective idealized
description, which would be obtained from a coarse-graining procedure \cite{Sasakura:2010rb}.  

The algebra \eq{eq:fuzzyalg} with \eq{eq:CeqM} and \eq{eq:GaussM},
\be
\phi_{x_1} \phi_{x_2} =M^G_{x_1x_2x_3} \phi_{x_3},
\ee
defines a fuzzy space of the kind satisfying \eq{eq:algcon1}, \eq{eq:algcon2} and \eq{eq:algcon3}. 
Intuitively, the function $\phi_x$ represents a fuzzy point at $x$, which
has fuzziness with a length scale $\sim 1/\sqrt{\beta}$.  
For the flat case $g_{ij}(x)=\delta_{ij}$, the Gaussian configuration \eq{eq:GaussM} becomes
\be
M^{flat}_{x_1x_2x_3}=B \exp\left[ -\beta \left((x_1-x_2)^2+(x_2-x_3)^2+(x_3-x_1)^2 
\right) \right].
\label{eq:GaussMflat}
\ee
This configuration indeed respects the Poincare symmetry and can be considered 
to represent a flat $D$-dimensional fuzzy space \cite{Sasai:2006ua}. 
In the limit $\beta\rightarrow \infty$ with 
an appropriate normalization $B$, the fuzzy space algebra approaches 
\be
\phi_{x_1}\phi_{x_2}=\delta^D(x_1-x_2)\, \phi_{x_1} \hbox{ in } \beta\rightarrow \infty.
\label{eq:flatalg} 
\ee
This is the limit to the usual continuous spaces with no fuzziness. Indeed the algebra 
\eq{eq:flatalg} is commutative and associative. In the rest 
of this paper, this limit will be denoted by the pointwise limit, and will be used 
to reproduce the constraint algebra of general relativity in Hamilton formalism from
tensor models. 

In fact, in the subsequent discussions,  
the Gaussian form \eq{eq:GaussM} or \eq{eq:GaussMflat} is not essential.
It is merely a representative of the configurations of $M_{abc}$, which have relatively 
local distributions when the $O(N)$ gauge symmetry \eq{eq:tensorsym} is appropriately fixed, and 
have a parameter that can be tuned to take the pointwise limit \eq{eq:flatalg}.
Another important thing is that  
one cannot take the pointwise limit as a starting point. As will be seen, one has to start with 
a finite fuzziness and then take the pointwise limit.
For simplicity, I will only use the flat expression \eq{eq:GaussMflat} in the subsequent computations,
but it should be straightforward to extend to the general cases by using the  
diffeomorphism invariant expression \eq{eq:GaussM}.  

\section{Realization of constraints by a vector model}
\label{sec:vector}
In this section, I will try to incorporate the constraint algebra of
general relativity \eq{eq:const1}, \eq{eq:const2} and 
\eq{eq:const3} in the framework of vector models in Hamilton formalism.
The discussions will proceed almost well, but some difficulties,
which will be discussed in Section \ref{sec:difficulty},
will arise in a final step.
Matrix models will also be abandoned due to the same difficulties.

The degrees of freedom of a vector model in Hamilton formalism are assumed to be given by 
$M_a\ (a=1,2,\ldots,N)$ 
and their conjugate momentums $\pi_a$. They are assumed to satisfy the canonical 
relation of Poisson bracket, 
\begin{equation}
\label{eq:vecbra}
\{ M_a, \pi_b\} =\delta_{ab}.
\end{equation}
Let me define
\begin{align}
\label{eq:vecham}
{\cal C}^V_{(ab)} &\equiv \frac12\left(\pi_a \pi_b +\epsilon' M_a M_b\right),\\
\label{eq:vecmom}
{\cal C}^V_{[ab]} &\equiv \frac12 \left(\pi_a M_b-\pi_b M_a\right), 
\end{align}
where $\epsilon'=\pm 1$ is a signature, and its relation with $\epsilon$ in 
\eq{eq:const1} or \eq{eq:grahh} 
will be given later.
The ${\cal C}^V$'s satisfy ${\cal C}^V_{(ab)}={\cal C}^V_{(ba)}$ and 
${\cal C}^V_{[ab]}=-{\cal C}^V_{[ba]}$, respectively. 
As will be explained in detail below, \eq{eq:vecham} and \eq{eq:vecmom} mimic 
the super-Hamiltonian and 
super-momentums of general relativity, respectively.

From \eq{eq:vecbra}, \eq{eq:vecham} and \eq{eq:vecmom}, one can straighforwardly 
obtain the following algebraic relations,
\begin{align}
\label{eq:vechh}
\{H_V(v^S),H_V(w^S)\}&=- \epsilon' D_V([v^S,w^S]),\\
\label{eq:vecdh}
\{D_V(v^A),H_V(w^S)\}&= H_V([v^A,w^S]), \\
\label{eq:vecdd}
\{D_V(v^A),D_V(w^A) \} &=  D_V([v^A,w^A]),
\end{align}
which look similar to \eq{eq:grahh}, \eq{eq:gradh} and \eq{eq:gradd} of general relativity.
Here $H_V$ and $D_V$ are defined by
\begin{align}
H_V(v^S)&\equiv v^S_{ab}\, {\cal C}^V_{(ab)}, \\
D_V(v^A)&\equiv v^A_{ab}\, {\cal C}^V_{[ab]},
\end{align}
and the upper indices $S$ and $A$ indicate the symmetric properties of the matrices,
$v^S_{ab}=v^S_{ba}$ and $v^A_{ab}=-v^A_{ab}$, respectively.
The $v,w$'s are assumed to be independent of the canonical variables.  
The square bracket $[v,w]$ denotes the commutator of matrices, $[v,w]_{ab}\equiv v_{ac}w_{cb}-w_{ac} v_{cb}$. 
Thus ${\cal C}^V_{(ab)}$ and ${\cal C}^V_{[ab]}$ form a Lie algebra
under the Poisson bracket. Especially, 
due to \eq{eq:vecdd} and the anti-symmetry of $v^A_{ab}$, 
${\cal C}^V_{[ab]}$ form the Lie algebra $so(N)$ of the orthogonal group, which 
is the kinematical symmetry of the vector model. 

In the following, I will show how the constraint algebra of general relativity 
\eq{eq:grahh}, \eq{eq:gradh} and
\eq{eq:gradd} can be reproduced from the algebra of a vector model \eq{eq:vechh}, \eq{eq:vecdh}
and \eq{eq:vecdd}.
Let me assume that a situation similar to the Gaussian configurations in Section \ref{sec:reviewtensor}
is occurring in the vector model.
Then the indices are assumed to take the coordinates of a continuous $D$-dimensional
space as in \eq{eq:contx}.

Let me first discuss \eq{eq:vecdd}.
In \eq{eq:vecdd}, the computation of the Poisson bracket has been reduced to
a commutator of matrices. Let me consider a (infinite-dimensional) matrix in the form,
\begin{align}
\label{eq:vecdefvaxy}
v^A_{xy}=\frac12(v^i(x)+v^i(y))\delta_i(x,y), 
\end{align}
which is anti-symmetric $v^A_{xy}=-v^A_{yx}$.
Here $v^i(x)$ are assumed to be arbitrary smooth functions on the $D$-dimensional space. 
Then the commutator between two such matrices is given by
\begin{align}
[v^A,w^A]_{xy}&=\frac14 \int dz\, (v^i(x)+v^i(z))\delta_i(x,z) (w^j(z)+w^j(y))\delta_j(z,y)-(x \leftrightarrow y).
\label{eq:veccomstart}
\end{align}
Because of the derivatives of the delta functions, the computation of the
right-hand side of  \eq{eq:veccomstart}
tends to become cumbersome. It is much more straightforward and easier to
do the computation by considering a test function $f(y)$.
By multiplying the right-hand side of \eq{eq:veccomstart} with $f(y)$ and 
performing the partial integrations over $y$ and $z$, one obtains
\begin{align}
&\int dz dy \, f(y) \left[(v^i(x)+v^i(z))\delta_i(x,z) (w^j(z)+w^j(y))\delta_j(z,y)-(x \leftrightarrow y)\right]
\nonumber \\
&\ \ =
2 v^i(x) (\partial_i \partial_j w^j(x)) f(x) +4 v^i(x)(\partial_i w^j(x)) \partial_jf(x) -(v\leftrightarrow w).
\label{eq:vecrescomstart}
\end{align} 
Comparing the last expression of \eq{eq:vecrescomstart} with the right-hand side of 
\begin{align}
\label{eq:compf} 
\int dy\, (v^i(x)+v^i(y))\delta_i(x,y)f(y)=(\partial_i v^i(x))f(x)+2 v^i(x) \partial_i f(x),
\end{align}
one obtains 
\begin{align}
[v^A,w^A]_{xy}=\frac12 \left([v,w]^i(x)+[v,w]^i(y)\right)\delta_i(x,y),
\label{eq:vecvawaxyres}
\end{align}
where
\begin{align}
[v,w]^i(x)\equiv v^j(x)\partial_j w^i(x)-w^j(x) \partial_j v^i(x).
\label{eq:vecdefvwi}
\end{align}
Comparing \eq{eq:vecvawaxyres} and \eq{eq:vecdefvwi} with \eq{eq:vecdefvaxy}, one concludes that
\eq{eq:vecdd} with \eq{eq:vecdefvaxy} exactly reproduces \eq{eq:gradd}. 
Thus $D_V(v^A)$ is the analogue to 
the generators of spatial diffeomorphism in the general relativity.

In the next place, let me discuss \eq{eq:vecdh}. Consider
\be
\label{eq:vecdefvsxy}
w^S_{xy}=c(\beta)\,w\left(\frac{x+y}2 \right) \exp(-\beta(x-y)^2),
\ee
where $(x-y)^2=(x-y)^i(x-y)^i$. The \eq{eq:vecdefvsxy} satisfies the symmetry $w^S_{xy}=w^S_{yx}$.
The function $w(x)$ is assumed to be an arbitrary smooth function on the $D$-dimensional space and 
its argument in \eq{eq:vecdefvsxy} takes the middle point between $x$ and $y$.
The coefficient $c(\beta)$ depending on $\beta$ will be determined later.

The Gaussian form in \eq{eq:vecdefvsxy} follows the Gaussian configurations 
in Section \ref{sec:reviewtensor}, and, 
at the final step of the following computations, 
the pointwise limit $\beta\rightarrow \infty$ will be taken to compare with
the constraint algebra of general relativity.
This Gaussian form is considered just because of its simplicity.
In fact, the following discussions do not depend on the details of the form.
What is needed is that $w^S_{xy}$ has distributions within finite ranges of 
relative distances between $x$ and $y$, and one can finally  
take a smooth pointwise limit.

The computation of the Poisson bracket \eq{eq:vecdh} has been reduced to 
the commutator of the matrices \eq{eq:vecdefvaxy} and \eq{eq:vecdefvsxy}, which is given by
\begin{align}
[v^A,w^S]_{xy}&=\frac{c(\beta)}{2}\int dz\,
(v^i(x)+v^i(z))\delta_i(x,z)\, w\left(\frac{z+y}2 \right) \exp(-\beta(z-y)^2)+(x\leftrightarrow y)\nonumber \\
&=\frac{c(\beta)}{2}\left[
\partial_iv^i(x) w\left(\frac{x+y}{2}\right) \exp(-\beta(x-y)^2)\right.
\nonumber\\
&\ \ \ \ \ \ \ \ \ \ \ \ \ \ \ \ \ \ \ \ \ +\left.
2 v^i(x) \partial^x_i\left(
w\left(\frac{x+y}{2}\right) \exp(-\beta (x-y)^2)\right)
\right]+(x\leftrightarrow y).
\label{eq:vecvaws}
\end{align}
The last expression is rather confusing, because the derivative with respect to $x$ in the last line
produces a factor $\beta$, which makes the pointwise limit difficult to handle. 
It is again much easier to do the computations by considering
 a test function $f(y)$. 
By multiplying the right-hand side of \eq{eq:vecvaws} with $f(y)$ and 
integrating over $y$, one obtains, in the leading order of $1/\beta$,  
\begin{align}
&\int dy f(y)\frac{c(\beta)}{2}\left[
\partial_iv^i(x) w\left(\frac{x+y}{2}\right) \exp(-\beta(x-y)^2)\right.
\nonumber\\
&\ \ \ \ \ \ \ \ \ \ \ \ \ \ \ \ \ \ \ \ \ \ \ \ \ \ \ +\left.
2 v^i(x) \partial^x_i\left(
w\left(\frac{x+y}{2}\right) \exp(-\beta (x-y)^2)\right)
+(x\leftrightarrow y)\right] \nonumber \\
&\ \ \ \ \ =
c(\beta) \left(\int dz \exp(-\beta z^2)\right) \left[ v^i(x)\partial_i w(x)+O(\beta^{-1}) \right] f(x).   
\end{align}
This concludes   
\begin{align}
[v^A,w^S]_{xy}=c(\beta)\, v^i \partial_i w \left( \frac{x+y}{2} \right) \exp \left(-\beta (x-y)^2\right)
\label{eq:veccomvwfinal}
\end{align} 
in the leading order of $1/\beta$.
Thus, by comparing the right-hand side of \eq{eq:veccomvwfinal} with \eq{eq:vecdefvsxy},
\eq{eq:vecdh} with \eq{eq:vecdefvaxy} and \eq{eq:vecdefvsxy} exactly reproduces \eq{eq:gradh}
in the pointwise limit $\beta\rightarrow \infty$. 

Finally let me discuss \eq{eq:vechh}. What should be computed is the commutator between the matrices
in the form \eq{eq:vecdefvsxy}:
\begin{align}
[v^S,w^S]_{xy}=c(\beta)^2 \int dz \left[v\left(\frac{x+z}2 \right)
w\left(\frac{z+y}2\right)-(x\leftrightarrow y)\right] \exp\left(-\beta(x-z)^2-\beta(z-y)^2
\right).
\end{align}
To systematically do the computation, let me again consider a test function $f(y)$ and evaluate
\begin{align}
c(\beta)^2 \int dydz\, f(y)\left[v\left(\frac{x+z}2 \right)
w\left(\frac{z+y}2\right)-(x\leftrightarrow y)\right] \exp\left(-\beta(x-z)^2-\beta(z-y)^2
\right).
\end{align}
By a change of variables, $y=a+b+x,\ z=b+x$, the integration becomes
\begin{align}
&c(\beta)^2 \int da db\, f(a+b+x)\left[ v(x+b/2)w(a/2+b+x)-(v\leftrightarrow w)\right]
\exp(-\beta a^2-\beta b^2).
\label{eq:compexp} 
\end{align}
Because of the Gaussian damping factor for $a$ and $b$, the large $\beta$ limit of this integral 
can well be evaluated after Taylor expanding in $a$ and $b$ the integrand other than the exponential.
Then one obtains 
\begin{align}
&c(\beta)^2 \beta^{-D-1} \left(\frac{1}{D} \int dz z^2 \exp(-z^2)\right)
\left(\int dz \exp(-z^2)\right)\nonumber \\
&\ \ \ \ \ \ \ 
\times \left[
\partial_i f(x) \left(v(x) \partial_i w(x) -w(x) \partial_i v(x)\right) +\frac12 f(x) 
\left(v(x) \partial_i \partial_i w(x) -w(x) \partial_i \partial_i v(x)\right)\right].
\label{eq:rescompexp}
\end{align} 
Comparing with \eq{eq:compf}, this concludes 
\begin{align}
\label{eq:vecvswsres}
[v^S,w^S]_{xy}=c_1 c(\beta)^2 \beta^{-D-1} \frac12 \left( l_i(x)+l_i(y) \right) \delta_i(x,y),
\end{align}
where $c_1$ is a numerical factor, and
\be
l_i(x)=v(x)\partial_i w(x)-w(x)\partial_i v(x).
\ee

The ugly contraction of the indices in \eq{eq:vecvswsres} comes from the simplified assumption
$z^2=z^iz^i$ in \eq{eq:vecdefvsxy}. If a general metric is assumed as $z^2=g_{ij} z^i z^j$, 
one will have 
\be
\label{eq:vecgauss}
\int da \sqrt{g}\ a^i a^j \exp(-\beta a^2) \sim \beta^{-\frac{D}{2}-1} g^{ij},
\ee
and \eq{eq:vecvswsres} will contain $g^{ij}$.
It would be straightforward to do the computations in a full covariant fashion following 
the expressions in Section \ref{sec:reviewtensor}.

Thus, by taking 
\begin{align}
c(\beta)&=\beta^{\frac{D+1}{2}}c_1^{-\frac12},
\label{eq:veccbeta}\\
\epsilon'&=-\epsilon,
\label{eq:relepe}
\end{align}
one obtains \eq{eq:grahh} of general relativity from \eq{eq:vechh}. 

An important fact in the above computations is that the finiteness of the  
range $|x-y|^2\lesssim 1/\beta$ of the distribution 
in \eq{eq:vecdefvsxy} plays an essential role
in the derivation of \eq{eq:vecvswsres}, even though the pointwise limit $\beta\rightarrow\infty$
is finally taken. 
This is indicated by the extra factor $\beta^{-1}$ in \eq{eq:vecvswsres}.  
This is also the reason why the final expression depends on the inverse metric $g^{ij}$,
which appears also in the constraint algebra \eq{eq:grahh} of general relativity.
If the matrix \eq{eq:vecdefvsxy} were assumed to have a full diagonal expression  
like $\delta(x,y)$ for instance, \eq{eq:grahh} would not be reproduced.   

\section{Difficulties and a solution}
\label{sec:difficulty}
From the discussions in Section \ref{sec:vector}, 
the super-Hamiltonians in the vector model should be given by ${\cal C}^V_{(ab)}$ in \eq{eq:vecham}.
Then the generators of an infinitesimal local ``time" translation will be given by 
$\delta t_{(ab)} {\cal C}^V_{(ab)}$. However, in the regime discussed in Section \ref{sec:reviewtensor},
the indices label each ``point" in a space, and the ``time" is generally highly non-local,
since $x$ and $y$ in $\delta t_{(xy)}$ can freely take any values.
Therefore this ``time" is very different from the conventional notion of time in physics.
One may instead consider a ``time" in a diagonal form $\delta t_{ab}\sim \delta t_a \delta_{ab}$.
But, as discussed in the last paragraph in Section \ref{sec:vector}, such a full diagonal form
cannot correctly reproduce the constraint algebra of general relativity.

There exists also another more general and serious difficulty. Under the Poisson bracket, 
the constraints \eq{eq:vecham} and \eq{eq:vecmom} form a Lie algebra with {\it constant} structure
constants. This means that the future time evolution is completely determined by the action of
the Lie group element parameterized by the ``time". As reviewed in Section \ref{sec:gr}, 
this is substantially different from the gravitational case, 
in which the constraint algebra has structure functions depending
on the metric. One cannot expect non-trivial
dynamics to occur from the constraint algebra of the vector model.  

These problems cannot be solved by considering a matrix model. One would try a set of constraints,
\begin{align}
{\cal C}^M_{(ab)}&=\pi_{ac}\pi_{bc}+\epsilon'M_{ac}M_{bc},\\
{\cal C}^M_{[ab]}&=\pi_{ac}M_{bc}-M_{ac}\pi_{bc}.
\end{align}
But the problems above appear in the same as in the vector model.

A solution can be given, if there exists a three-index tensor $M_{abc}$. 
By using $M_{abc}$, the two indices of ${\cal C}^V_{(ab)}$ can be contracted as
\be
\label{eq:calcva}
{\cal C}^V_a\equiv M_{abc} {\cal C}^V_{(ab)}.
\ee 
Then the infinitesimal time can have only one index as $\delta t_a {\cal C}^V_a$.
Moreover, in the regime explained in Section \ref{sec:reviewtensor}, 
$\delta t_x M_{xyz}$ will provide a 
non-diagonal distribution of a finite range for $y,z$, that is necessary to reproduce the constraint 
algebra of general relativity in the pointwise limit $\beta\rightarrow \infty$, as
is discussed in the last paragraph of Section \ref{sec:vector}.
And also, as will be seen in Section \ref{sec:tensor}, the constraint algebra
does not have structure constants, but they rather depend on $M_{abc}$ and its conjugate. 

In principle, the indices of the super-momentums of the vector model ${\cal C}^V_{[ab]}$ 
can also be contracted by $M_{abc}$. However, this is not a valid option, because,
as explained in Section \ref{sec:reviewtensor}, the tensor models have the orthogonal 
group symmetry, and ${\cal C}^V_{[ab]}$ are the generators of this kinematical symmetry. 
If they were contracted, the number of constraints would be reduced, and
the symmetry could not be fully incorporated by the constraints. 

Another important thing is that $M_{abc}$ must be a dynamical variable.
If not, the Poisson bracket between the super-momentums ${\cal C}^V_{[ab]}$ and ${\cal C}^V_a$
would not close: 
\begin{align}
\{v^A_{ab} {\cal C}^V_{[ab]},v^c M_{cde} {\cal C}^V_{(de)}\}
&=v^A_{ab} v^c M_{cde} \{{\cal C}^V_{[ab]},{\cal C}^V_{(de)}\}\\
&=[v^A, vM]_{ab} {\cal C}^V_{(ab)},
\end{align}
where $vM$ denotes a matrix $vM_{ab}\equiv v^cM_{cab}$. 
The expression in the last line does not in general have the form of 
linear combinations of \eq{eq:calcva}.
Therefore one has to include $M_{abc}$ as a canonical variable, and 
make it transform appropriately under the Poisson bracket with the super-momentums.

The discussions in this section can be summarized as follows. 
To correctly reproduce the constraint algebra of general relativity and the usual notion of time, 
it is necessary to include a three-index tensor $M_{abc}$ as a dynamical variable. 
In this sense, $M_{abc}$ is the dynamical variable corresponding to gravity.
The vector and matrix variables may be incorporated consistently in the constraint algebra,
but they are not necessary. Instead they can rather be regarded as some matter degrees of freedom,
which can be added consistently, as will be discussed in Section \ref{sec:matter}.

\section{Realization by a rank-three tensor model}
\label{sec:tensor}
The discussions in Section \ref{sec:difficulty} imply that the pure gravitational 
system should be obtained from a rank-three tensor model which has a three-index tensor as
its only dynamical variable. 
The canonical variables are assumed to be given by $M_{abc}$ and $\pi_{abc}$.
They are assumed to satisfy the generalized Hermiticity condition
\eq{eq:tengenher1} and its conjugate correspondence, 
\be
\pi_{abc}=\pi_{bca}=\pi_{cab}=\pi^*_{bac}=\pi^*_{acb}=\pi^*_{cba},
\label{eq:tengenher2}
\ee
respectively.
The Poisson bracket between the canonical variables is assumed to be given by
\be
\{M_{abc},\pi_{def}\}=\delta_{abc,def}\equiv \delta_{ad}\delta_{be}\delta_{cf}+
\delta_{ae}\delta_{bf}\delta_{cd}+\delta_{af}\delta_{bd}\delta_{ce}.
\ee

Let me consider  
\begin{align}
{\cal C}_{(ab)}&\equiv \frac{1}{2}\left(\pi_{acd} \pi_{bdc} -\epsilon M_{acd} M_{bdc}\right), 
\label{eq:tendefcsym}\\
{\cal C}_{[ab]}&\equiv \frac{1}{2}\left(\pi_{acd} M_{bcd}-M_{acd} \pi_{bcd} \right),
\label{eq:tendefcanti}  
\end{align}
where the relation $\epsilon'=-\epsilon$ in \eq{eq:relepe} has already been used. 
Let me define 
\begin{align}
H'(v^S)&\equiv v^S_{ab} {\cal C}_{(ab)},\\
D(v^A)&\equiv v^A_{ab} {\cal C}_{[ab]},
\end{align}
where $v^S_{ab}=v^S_{ba}$ and $v^A_{ab}=-v^A_{ba}$, and they are assumed to be
independent of the canonical variables.
Then $H'$ and $D$ satisfy
\begin{align}
\{H'(v^S),H'(w^S)\}&=\epsilon D([v^S,w^S]),
\label{eq:tenhh}\\
\{D(v^A),H'(w^S)\}&=H'([v^A,w^S]),\label{eq:tendh}\\
\{D(v^A),D(w^A)\}&=D([v^A,w^A]),\label{eq:tendd}
\end{align}
which are actually the same as \eq{eq:vechh}, \eq{eq:vecdh} and \eq{eq:vecdd}
of the vector model.

The discussions in Section \ref{sec:vector} for a vector model uses only the fact that 
the algebra of constraints under Poisson bracket can be written as commutators of matrices as in 
\eq{eq:vechh}, \eq{eq:vecdh} and \eq{eq:vecdd}. Since this is the same for the tensor model
as in \eq{eq:tenhh}, \eq{eq:tendh} and \eq{eq:tendd}, the constraint algebra of general relativity 
can be reproduced in the same way as in the vector model.

As discussed in Section \ref{sec:difficulty}, let me contract the indices of ${\cal C}_{(ab)}$
with $M_{abc}$ to construct the possible super-Hamiltonians defined by  
\be
\bar{\cal C}_a\equiv M_{abc} ({\cal C}_{(bc)}+\lambda \delta_{bc}),
\label{eq:tendefcbar}
\ee
where I have included a new term with a coefficient $\lambda$.
This term is meaningful because it adds a non-constant term to $\bar{\cal C}_a$.
The term is consistent with the kinematical symmetry of the tensor models, since 
$\delta_{ab}$ is an invariant of $O(N)$.  

Now let me define
\be
\label{eq:tendefbarh}
\bar H (v)\equiv v_a \bar {\cal C}_a,
\ee
for an infinitesimal parameter $v_a$ independent of the canonical variables.
Because of the $O(N)$ invariant form of \eq{eq:tendefbarh}, it is obvious that
\be
\{D(v^A),\bar H(w)\} =\bar H(v^Aw),
\ee
where $v^Aw_a\equiv v^A_{ab}w_b$, and therefore 
the Poisson bracket between $D$ and $\bar H$ closes. 
 
The Poisson bracket between two $\bar H$'s is given by
\begin{align}
\label{eq:poihh}
\{\bar H(v), \bar H(w)\}&=
\{\bar{\cal C}_{ab},\bar{\cal C}_{cd}\}\bar v_{ab}\bar w_{cd}
+\{\bar v_{ab},\bar{\cal C}_{cd}\}\bar w_{cd} \bar{\cal C}_{ab}
+\{\bar{\cal C}_{ab},\bar w_{cd}\}\bar v_{ab} \bar{\cal C}_{cd}
+\{\bar v_{ab},\bar w_{cd}\}\bar{\cal C}_{ab} \bar{\cal C}_{cd},
\end{align}
where I have defined
\begin{align}
\bar{\cal C}_{ab}&\equiv {\cal C}_{(ab)}+\lambda \delta_{ab},\\
\bar v_{ab}&\equiv v_c M_{(ab)c},\label{eq:tenvab}\\ 
\bar w_{ab}&\equiv w_c M_{(ab)c},\label{eq:tenwab}
\end{align}
for notational simplicity. Here $M_{(ab)c}$ (and $\pi_{(ab)c}$) is the symmetrization defined by  
\begin{align}
M_{(ab)c}&\equiv \frac12\left(M_{abc}+M_{bac}\right),\\
\pi_{(ab)c} &\equiv \frac12 \left( \pi_{abc}+\pi_{bac} \right).
\end{align}
Note that, thanks to \eq{eq:tengenher1} and \eq{eq:tengenher2}, $M_{(ab)c}$ and $\pi_{(ab)c}$
are totally symmetric with respect to their indices.
The computation of the first term in \eq{eq:poihh} is similar to \eq{eq:tenhh}. The last
term in \eq{eq:poihh} trivially vanishes. The sum of the second and the third terms can be computed as
\begin{align}
&\frac{1}{2} v_e \{ M_{(ab)e},\pi_{cfg}\pi_{dgf}-\epsilon M_{cfg}M_{dgf} \}
w_h M_{(cd)h} \bar {\cal C}_{ab}-(v\leftrightarrow w) \nonumber \\
&\ \ = v_e \pi_{cfg} \delta_{(ab)e,dgf}w_h M_{(cd)h} \bar {\cal C}_{ab}-(v\leftrightarrow w)\nonumber\\
&\ \ =2  \, (v_a w_b-w_a v_b)\, \bar{\cal C}_{cd} \, \pi_{(ac)e}\, M_{(bd)e}. 
\end{align}
Thus the result of \eq{eq:poihh} is obtained as 
\be
\label{eq:tenresbarhbarh}
\{\bar H(v), \bar H(w)\}=\epsilon D([\bar v,\bar w])+
2 (v_a w_b-w_a v_b)\, \bar{\cal C}_{cd} \, \pi_{(ac)e}\, M_{(bd)e}.
\ee 
The first term of the right-hand side is what is desired. 
The matrices $\bar v$ and $\bar w$ are distributed by $M_{abc}$ as in \eq{eq:tenvab} and \eq{eq:tenwab}.
Therefore, this term exactly reproduces 
the constraint algebra of general relativity in the pointwise limit $\beta\rightarrow \infty$,
as has been discussed in Section \ref{sec:vector}.
However, the second term is proportional to neither ${\cal C}_{[ab]}$ 
nor $\bar {\cal C}_a$. 
Therefore $\bar {\cal C}_a$ are not appropriate as super-Hamiltonians, since the constraint
algebra does not close.

To cancel the second term in \eq{eq:tenresbarhbarh}, let me 
first proceed under the assumption that the super-Hamiltonians be invariant under the time-reversal
transformation $\pi_{abc}\rightarrow -\pi_{abc}$. I also assume that the terms are at most cubic
and local: each term must be connected, not like $M_{abb}M_{cde}M_{cde}$ for instance.
Then the possible cubic terms other than those in \eq{eq:tendefcbar} can be listed as
\begin{align}
{\cal C}^{(1)}_a &\equiv \pi_{abc} \pi_{bde} M_{cde},\\
{\cal C}^{(2)}_a &\equiv M_{abc} \pi_{bcd} \pi_{dee},\\
{\cal C}^{(3)}_a &\equiv \pi_{abc} M_{bcd} \pi_{dee},\\
{\cal C}^{(4)}_a &\equiv \pi_{abc} \pi_{bcd} M_{dee},\\
{\cal C}^{(5)}_a &\equiv M_{abc} M_{bcd} M_{dee},
\end{align}
where I have not cared the order of the indices, since this is not 
essential in the following discussions. It can easily be shown that 
${\cal C}^{(2)},\ldots,{\cal C}^{(5)}$ are not useful to 
cancel the second term in \eq{eq:tenresbarhbarh}. 
When ${\cal C}^{(1)}$ is added to \eq{eq:tendefcbar},
the Poisson bracket will have additional contributions, $\{ {\cal C}^{(1)}, {\cal C}^{(1)}\}$
and $\{ {\cal C}^{(1)},\bar {\cal C}\}$. From explicit computations, 
$\{ {\cal C}^{(1)},{\cal C}^{(1)}\}$ is proportional to super-momentums, 
and might be allowed. However, $\{ {\cal C}^{(1)},\bar {\cal C}\}$ 
contains terms
\be
v_a w_b \{ {\cal C}_a^{(1)},\bar {\cal C}_b\} \sim d_1\,v_a w_e M_{def} \pi_{acd}
M_{fij} M_{cji} + d_2\,
v_a w_b M_{def}M_{bge} M_{gcf}\pi_{acd}+\cdots,
\label{eq:tenc1c}
\ee
where $d_i$'s are some integers.
The first term in \eq{eq:tenc1c} can be used to cancel part of the unwanted term 
in \eq{eq:tenresbarhbarh}, but 
the second term in \eq{eq:tenc1c} is problematic. This term is not proportional to the 
super-momentums, and it can also be shown that this term cannot be canceled by adding  
${\cal C}^{(2)},\ldots,{\cal C}^{(5)}$.  
Thus it is not possible to construct super-Hamiltonians
that contain \eq{eq:tendefcbar}, are cubic at highest and local,
and respect the time-reversal symmetry. 

A simple solution can be found, if one relaxes the time-reversal symmetry, while 
the other assumptions are kept intact. 
Let me consider 
\be
{\cal C}_a\equiv( M_{abc}+ \epsilon'' \pi_{abc})({\cal C}_{(bc)}+\lambda \delta_{bc}),
\label{eq:tendefc}
\ee
where $\epsilon''$ is a parameter which will be determined in the following.
This will become the final form of super-Hamiltonians.
Define
\begin{align}
H(v)=v_a {\cal C}_a
\label{eq:tendefh}
\end{align}
for an infinitesimal parameter $v_a$ independent of the canonical variables.
Because of the $O(N)$ invariant form of \eq{eq:tendefh}, one obtains
\be
\{D(v^A),H(w)\} =H(v^Aw),
\label{eq:finalDH}
\ee
where $v^Aw_a\equiv v^A_{ab}w_b$.

The Poisson bracket between $H$'s is given by
\begin{align}
\label{eq:poihhfinal}
\{ H(v), H(w)\}&=
\{\bar{\cal C}_{ab},\bar{\cal C}_{cd}\}v'_{ab}w'_{cd}
+\{v'_{ab},\bar{\cal C}_{cd}\}w'_{cd} \bar{\cal C}_{ab}
+\{\bar{\cal C}_{ab},w'_{cd}\}v'_{ab} \bar{\cal C}_{cd}
+\{v'_{ab},w'_{cd}\}\bar{\cal C}_{ab} \bar{\cal C}_{cd},
\end{align}
where 
\begin{align}
v'_{ab}&\equiv v_c (M_{(ab)c}+\epsilon'' \pi_{(ab)c}),\label{eq:tenvpab}\\ 
w'_{ab}&\equiv w_c (M_{(ab)c}+\epsilon'' \pi_{(ab)c})\label{eq:tenwpab}.
\end{align}
Computations similar to the case of \eq{eq:poihh} result in 
\be
\label{eq:tenreshh}
\{H(v), H(w)\}=\epsilon D([v',w'])+
2 \left(1-\epsilon (\epsilon'')^2 \right)(v_a w_b-w_a v_b)\, 
\bar{\cal C}_{cd} \, \pi_{(ac)e}\, M_{(bd)e}.
\ee 
For the closure of the constraint algebra, $\epsilon''$ is determined as
\be
\epsilon''=\pm1/\sqrt{\epsilon}.
\label{eq:valepp}
\ee
Then 
\be
\{H(v), H(w)\}=\epsilon D([v',w']).
\label{eq:finalHH}
\ee

The commutator in \eq{eq:finalHH} is of the matrices \eq{eq:tenvpab} and 
\eq{eq:tenwpab}.  
This contains not only $M_{abc}$ but also $\pi_{abc}$.
But, as is stressed in Section \ref{sec:vector}, the constraint algebra of 
general relativity can be reproduced without the details of the distribution of these matrices.
One would be able to expect that,
in the regime discussed in Section \ref{sec:reviewtensor},
 not only $M_{abc}$ but also the conjugate momenta $\pi_{abc}$
have relatively local distributions which become pointwise in the limit $\beta\rightarrow\infty$.
Therefore the addition of $\pi_{abc}$ will not affect the derivation of the constraint algebra of
general relativity in the pointwise limit.

In the Euclidean case $\epsilon=-1$, \eq{eq:valepp} implies that 
$\epsilon''$ must take an imaginary value $\epsilon''=\pm i$. 
Then an infinitesimal time-like shift will generate
\be
\delta M_{abc}=\{ M_{abc},H(v)\}=\pm i v_a \bar {\cal C}_{bc} +\cdots.
\ee
This direct appearance of the imaginary unit $i$ will generally violate 
the generalized Hermiticity condition \eq{eq:tengenher1}. 
Therefore the Euclidean case is not consistent. 

\section{Coupling with matters}
\label{sec:matter}
Although the vector model considered in Section \ref{sec:vector} is not suited for describing 
gravity, it may be added as a matter. Because of the kinematical character,
the total super-momentums are simply given by the sum of those of the 
three-tensor \eq{eq:tendefcanti} and the vector \eq{eq:vecmom} as
\be
{\cal C}_{[ab]}^{total}\equiv{\cal C}_{[ab]}+{\cal C}^V_{[ab]},
\label{eq:mattermom}
\ee
which are obviously the generators of the $O(N)$ transformation.
The total super-Hamiltonians would be defined by a summation, 
\be
{\cal C}_a^{total} \equiv {\cal C}_a+(M_{abc}+\epsilon'' \pi_{abc}){\cal C}^V_{(bc)},
\label{eq:matterham}
\ee
where ${\cal C}_a$ and ${\cal C}^V_{(ab)}$ are the super-Hamiltonians of 
the three-tensor \eq{eq:tendefc} and the vector \eq{eq:vecham}, respectively.  
Here the coupling between gravity and matter is
realized by the contraction of ${\cal C}^V_{(ab)}$ with $M_{abc}+\epsilon'' \pi_{abc}$.
To check whether this actually produces the 
desired result, let me compute the Poisson bracket,
\begin{align}
\{ H^{total}(v),H^{total}(w) \}=v_a w_b \left(
\{{\cal C}_a,{\cal C}_b \}
+\{{\cal C}_a,M_{bcd}+\epsilon''\pi_{bcd}\} {\cal C}^V_{(cd)}
+\{M_{acd}+\epsilon''\pi_{acd},{\cal C}_b\} {\cal C}^V_{(cd)}\right.
\nonumber\\ 
\left.
+\{M_{acd}+\epsilon''\pi_{acd},M_{bef}+\epsilon''\pi_{aef}\}
{\cal C}^V_{(cd)}{\cal C}^V_{(ef)} 
+(M_{acd}+\epsilon''\pi_{acd})(M_{bef}+\epsilon''\pi_{bef})
\{{\cal C}^V_{(cd)},{\cal C}^V_{(ef)}\} 
\right),
\label{eq:htothtot}
\end{align}
where 
\be
H^{total}(v)\equiv v_a {\cal C}_a^{total}.
\ee
The first and the last terms in \eq{eq:htothtot} produce the desired form, and the fourth term
obviously vanishes. The computation of the sum of the second and the third terms 
is essentially the same as \eq{eq:poihhfinal}, and one obtains
\be
\label{eq:totreshh}
\{H^{total}(v), H^{total}(w)\}=\epsilon D^{total}([v',w']),
\ee
if the condition \eq{eq:valepp} is satisfied, where
\be
D^{total}(v^A)=v^A_{ab}{\cal C}^{total}_{[ab]},
\ee
and $v'$ and $w'$ are defined in \eq{eq:tenvpab} and \eq{eq:tenwpab}, respectively.
Because of the apparent $O(N)$ invariant forms, it is straightforward to derive
\be
\{D^{total}(v^A),H^{total}(w)\} =H^{total}(v^Aw).
\label{eq:finaltotalDH}
\ee

It is obvious that one can add various matters in similar manners.
It will be straightforward to extend \eq{eq:vecbra}, \eq{eq:vecham}, \eq{eq:vecmom}, 
\eq{eq:mattermom} and \eq{eq:matterham} to such general cases.
Any rank of tensors and statistics, bosonic or fermionic, will be allowed.

\section{Summary, discussions and future prospects}
\label{sec:sum}
In this paper, I have discussed how {\it local} time can be introduced in tensor models.
Since it was not clear how to formulate such tensor models in the Lagrangian formalism,  
a rank-three tensor model with first class constraints in Hamilton formalism 
has been presented.
The discussions have made it clear that a three-index tensor has 
a prominent feature necessary for the purpose.
The other rank tensors can be added as matter sectors which has coupling with the three-index tensor.
In this sense, the rank-three tensor model, which contains a three-index tensor as its only
dynamical variable, can be regarded as the gravitational sector 
which can universally couple with matters.

The momentum constraints have straightforwardly been constructed so that
they incorporate the kinematical symmetry of the rank-three tensor models.
Then the consistency of the local time evolution 
requires Hamiltonian constraints and the momentum constraints to compose
a closed first class constraint algebra.
This closure condition gives strong limitations on the possible forms of super-Hamiltonians,
and they have been determined on the assumptions that they are local and cubic at highest.
They have turned out to contain terms which break the time-reversal symmetry.
It has also been shown that the constraint algebra reproduces the first class constraint
algebra of general relativity in the limit of usual continuous spaces. 
The first class constraint algebra closes exactly without any approximations,   
and this is a good feature which marks distinction from the previous approaches \cite{Piran:1985ke,Friedman:1986uh,
Gambini:2005vn,Dittrich:2011ke,Bahr:2011xs} to discrete gravity in the canonical formalism.

The most important implication of this paper is that 
it is in principle possible for the rank-three tensor model
to have the first class constraint algebra which reproduces that of general relativity
in the limit of usual continuous spaces. 
However, the limiting procedure  is rather formal
assuming the regime discussed in Section \ref{sec:reviewtensor}. 
Therefore it will be necessary to investigate whether the limit 
can be generated by the actual dynamics of the rank-three tensor model.
Not only the constraint algebra, but it would also be necessary to check 
whether the equations of motion of general relativity can directly be reproduced
in the continuum limit from the super-Hamiltonians.  
Especially, the fate of the terms with the breaking of the time-reversal symmetry 
has to be studied. If the breaking effect has turned out to be too large to be allowed 
by experiments/observations, the super-Hamiltonians must be reconsidered
under relaxed assumptions. By canonical transformations of the variables, it is also possible to change the property of the super-Hamiltonians under the time-reversal transformation. 

Regardless of whether such a regime in Section \ref{sec:reviewtensor} actually exists in the dynamics,
the canonical rank-three tensor model of this paper has a significance of its own 
as a model of emergent space and gravity with consistently incorporated local time.
The form of the super-Hamiltonians has similarity with the Hamiltonian of the $c<1$ string 
field theory \cite{Ishibashi:1993pc,Ikehara:1994vx,Ambjorn:2008ta},
which has applications to random surfaces.
Since the $c<1$ string field theory may be regarded as (1+1)-dimensional gravity with matters,
the rank-three tensor model of this paper may have some connections to it.
The connections may provide an interesting testing ground as well as 
some hints for the dynamics of the tensor model.


\bibliography{geotensor}

\end{document}